\documentclass[prl,preprint,showpacs,preprintnumbers,amsmath,amssymb,floatfix]{revtex4}

\usepackage{amsthm}
\usepackage{graphicx}
\usepackage{dcolumn}
\usepackage{bm}
\usepackage{verbatim}

\begin{document}

\bibliographystyle{plainnat}
\title{Negative specific heat in a quasi-2D generalized vorticity model}
\author{T. D. Andersen}
\email{andert@rpi.edu}
\author{C. C. Lim}
\email{limc@rpi.edu}
\affiliation{Mathematical Sciences, RPI, 110 8th St., Troy, NY, 12180}
\date{\today}

\begin{abstract}
Negative specific heat is a dramatic phenomenon where processes decrease in temperature
when adding energy.  It has been observed in gravo-thermal collapse of
globular clusters.  We now report finding this phenomenon in bundles of nearly parallel,
 periodic, single-sign generalized vortex filaments in the electron magnetohydrodynamic (EMH) model for the unbounded plane under strong magnetic
 confinement.  We derive the specific heat using a steepest descent method and a mean field property. Our derivations show that as temperature increases, the overall size of the system increases
 exponentially and the energy drops.  The implication of negative specific heat is a runaway
 reaction, resulting in a collapsing inner core surrounded by an expanding halo of filaments.
\end{abstract}

\maketitle
Negative specific heat is an unusual phenomenon first discovered in 1968 in microcanonical (isolated system) statistical equilibrium models of
gravo-thermal collapse in globular clusters \cite{Lynden:1968}.  In gravo-thermal collapse, a disordered system of stars in isolation under-goes a process of core collapse with the following steps: (1) faster stars are lost to an outer halo where they slow down, (2) the loss of potential (gravitational) energy causes the core of stars to collapse inward some small amount, (3) the resulting collapse causes the stars in the core to speed up.  If one considered the ``temperature'' of the cluster to be the average speed of the stars, this process has negative specific heat because a loss of energy results in an increase in overall temperature.  Although one does not need to consider the cluster as a collection of molecules with negative specific heat to study its behavior, it can lead to insight into similar system, which has led us to this letter's findings.

In the intervening four decades, negative specific heat has been
observed in few other places.  Its counterintuitive implication is that as a system
loses energy its temperature increases and as one adds energy the system cools.  In globular clusters, this results in a runaway reaction leading to a collapse of the cluster.  In a
magnetic fusion system or other thermally isolated plasma, should negative specific heat exist, the related runaway collapse could have profound implications for fusion where extreme confinement is critical to a sustained reaction.

While \cite{Schrodinger:1952} has proven that systems that are not isolated from the environment must have positive specific heat, 
the specific heat in isolated systems can be negative \cite{Lynden:1977}.  We now report finding negative specific heat in the thermally microcanonical system of confined generalized vorticity filaments, a finding of great interest to magnetohydrodynamics (MHD) and plasma physics.

Our results have general applicability to vortex systems, and, although the calculations here-in presented have not yet appeared elsewhere, we have chosen a specific application to illustrate the value of our approach.  In this letter, we address a plasma model known as the electron magnetohydrodynamical (EMH) model.  

Typically, magnetohydrodynamic plasma models are two fluid models, requiring equations governing the electron motion and equations governing the ion motion coupled together \cite{Uby:1995}.  The EMH model bypasses the two-fluid model by representing the electron fluid and the magnetic field as a single, generalized fluid with a neutralizing ion background that is stationary on the timescale chosen.  

The EMH model takes the magnetic field, ${\bf B} = \nabla\times {\bf A}$, and the charged fluid vorticity, ${\bf \omega} = \nabla\times{\bf v}$, and combines them into a general vorticity field $\Omega = \nabla\times {\bf p}$ where the generalized momentum, ${\bf p} = m{\bf v} - e{\bf A}$, $m$ is the electron mass, $-e$ is the electron charge, ${\bf v}$ is the fluid velocity field, and ${\bf A}$ is the magnetic vector potential field.  For a brief overview of the model, see \cite{Uby:1995}.  A detailed model discussion can be found in \cite{Gordeev:1994}.

Our goal is to find the specific heat of this vortex model in statistical equilibrium given an appropriate definition for energy and a microcanonical  probability distribution.  Our approach is to describe the statistical behavior of a system of discrete, interacting 
vortex structures, rather than continuous vorticity, and consider the limiting case of an infinite number of infinitesimally weak 
vortex structures
recovering the continuous vorticity field.

The first step in our approach is to modify the continuous vorticity field, $\Omega$, such that it describes a large number, $N$, of discrete vortex filaments, continuous in one dimension but point-like in the other two.  Therefore, we assume that, instead of being continuously varying throughout, $\Omega$ has a large number of periodic filaments that are nearly parallel to the z-axis.  The nearly parallel assumption comes from the assumption of strong magnetic confinement and conservation of angular momentum which polarizes the filaments.  The period, $L$, is assumed to be unity, $L=1$, without loss of generality since all other distances and distance-dependent quantities can be scaled by $L$.  

The vorticity field, which depends on space, ${\bf r}\in\Re^3$, now looks like:
\begin{equation}
\Omega({\bf r}) = \sum_{i=1}^N\int_0^1 d\tau\Gamma_i\delta({\bf r} - {\bf r}_i(\tau)),
\label{eqn:vort}
\end{equation} where ${\bf r}=(x,y,z)$ and ${\bf r}_i=(x_i,y_i,z_i)$.  Periodicity requires that ${\bf r}_i(0)={\bf r}_i(1)$.  Therefore, we assume that the discretization of ${\bf r}$ into $\sum_i {\bf r}_i$ has this property in Equation \ref{eqn:vort}.  Because of the nearly parallel constraint, the arclength, $\tau$, has the property that $\tau\sim z_i\forall i$.  For simplicity, we will represent ${\bf r}_i(\tau)=(x_i,y_i,\tau)$ as a complex number $\psi_i(\tau) = x_i(\tau) + iy_i(\tau)$.  

To simulate the vortex system on a computer, one must use a discrete field, but, even in our analytical approach, it is easier to start with a finite number of filaments and take the limit, $N\rightarrow\infty$, later, keeping total vorticity, $\Lambda = \int_{\Re^3} \Omega({\bf r}) d{\bf r}$, constant by rescaling.  This is known as a {\bf non-extensive thermodynamic limit} approach because the overall vortex strength stays constant even as 
the number of vortex filaments increase towards infinity.  This common procedure is efficient in recovering the continuous vorticity field statistics.

We use an approximate model for vortex behavior known as the local-induction approximation (LIA), useful for nearly parallel filaments, combined with a two-dimensional logarithmic interaction.  Heuristically, this model comes from two separate results: One result, that of  \cite{Kinney:1993}, shows that a
discrete field of \emph{perfectly} parallel vortices has a logarithmic interaction.  Another result for a single filament in 3D, that of \cite{Uby:1995}, is a well-known first order approximation for the motion of fluid filaments, there-in extended to the EMH model.  This LIA approximation, derived from the Biot-Savart law of magnetic/velocity induction for filaments of magnetism/vorticity, depends on the binormal of the curve of the filament in 3-space causing Brownian variations.

The combination of the two results of \cite{Kinney:1993} and \cite{Uby:1995} yields the familiar London free energy of type-II superconductors \cite{Nordborg:1998}, equally valid for generalized vorticity \cite{Uby:1995}:
\begin{align}
E_N = &\alpha\int_0^1d\tau \sum_{i=1}^{N} \Gamma_i/2\left|\partial \psi_i(\tau)/\partial\tau
\right|^2 -\nonumber\\ &\int_0^1d\tau\sum_{i=1}^N\sum_{j>i}^N \Gamma_i\Gamma_j\log|\psi_i(\tau) - \psi_j(\tau)|,
\label{eqn:Ham}
\end{align} where $\alpha$ is the vortex elasticity in units of energy/length.
The generalized angular momentum (angular momentum plus magnetic moment), from now on simply refered to as the angular momentum, is
\begin{align}
M_N = \sum_{i=1}^N\Gamma_i\int_0^1 d\tau |\psi_i(\tau)|^2.
\label{eqn:angmo}
\end{align}   $\Gamma_i$ is the vortex circulation.  From this point on, we assume the vortex circulations are all scaled to unity, $\Gamma_i=1\forall i$, not entirely general but sufficient for the following results.  

Although we say the London free energy, we note that it is
not describing the same type of system that it would describe in the type-II 
superconductor setting because these vortex filaments are generalized and not purely flux lines.

What makes this system perhaps even more interesting is that it exists in an unbounded plane rather than with periodic boundary conditions, appropriate to superconductor research but not here, doing away with artificial planar periods.  Other sources for this vorticity description as well as others can be found in \cite{Neu:1990}, \cite{Chorin:1991}, \cite{Lim:2006}, and \cite{Klein:1995, Lions:2000}.  As far as we are aware, we are the first to apply it to the EMH model.

For our isolated, classical system, the energy and angular momentum plus magnetic moment are conserved giving rise to the following probability distribution for the filaments in equilibrium:
\begin{equation}
P(s) = Z^{-1}\delta(NH_0 - E_N - pM_N)\delta(NR^2 - M_N),
\label{eqn:gibbs}
\end{equation} where $H_0$ is the total ``enthalpy'' per vortex per period of the plasma, $s$ is the complete state of the system, and $Z=\int ds\delta(NH_0 - E_N - pM_N)\delta(NR^2-M_N)$ is a normalizing factor called the partition function.  Here $E_N$ is the energy functional and $M_N$ is the angular momentum.  It is our intent to allow
$R^2$, the size of the system,
\begin{equation}
 R^2 = \lim_{N\rightarrow\infty} \left\langle N^{-1}\int_0^1 d\tau |\psi_i(\tau)|^2\right\rangle,
\end{equation} to be determined by other parameters in the system and keep enthalpy and pressure, $p$, fixed.  This distribution fully describes the density distribution of the filaments in the plane for every parameter.

The size of the configuration space (partition function), $Z$, cannot be found in closed form by any known analytical methods.  Since our aim is an explicit expression for specific heat, we need to know $Z$ in closed form.

To simplify the equations, we combine the large number of vortices into two average or ``mean'' vortices, which results in a mean vorticity field.  This is the ``mean-field'' approach common in statistical mechanics.  We also exploit the clear axisymmetry of the system and reduce the space from three to two dimensions.  Given this approach, our mean vortices are as follows: one mean vortex is a mean distance from the origin.  The other is the statistical center of charge of all the filaments--a single, perfectly straight filament fixed at the origin with strength of the remaining vortices, $N-1\sim N$.  

Given a filament $i$ and a filament $j$, the mean-field approximation implies the following:
\begin{equation}
\langle|\psi_i - \psi_j|\rangle\longrightarrow\sqrt{\int_0^1 d\tau |\psi_i(\tau)|^2}=||\psi_i||,
\label{eqn:rsqDef}
\end{equation} where $i$ is any filament index and the double bars, $||\cdot||$, indicate $\mathcal{L}_2$-norm on the interval $[0,1]$.

The energy function now reads as follows:
\begin{align}
E_N' &= \int_0^1d\tau \sum_{i=1}^{N} \left[\frac{\alpha}{2}\left|\frac{\partial \psi_i}{\partial\tau}
\right|^2 - \frac{N}{4}\log ||\psi_i||^2\right].
\label{eqn:HamPrime}
\end{align}  This assumption changes the interaction between vortices from fully coupled to fully decoupled, that is, all vortices are statistically independent, and the statistics of
the entire system can be found from the statistical behavior of a single vortex.  We modify \ref{eqn:gibbs} and \ref{eqn:angmo} appropriately and, from here on forth, drop primes since neither the full energy nor the complete distribution nor the full angular momentum will be further referenced.

In statistical mechanics of isolated systems, all equilibrium statistics can be determined from maximizing the entropy.  The entropy per filament, $S_N$, is defined by,
\begin{equation}
 e^{S_N} = \int D\psi \delta(NH_0-E_N-pNR^2)\delta(NR^2-M_N).
 \label{eqn:eS}
\end{equation}
This definition implies,
\begin{equation}
 S_N = \log\left[\int D\psi \delta(NH_0-E_N-pNR^2)\delta(NR^2-M_N)\right].
\end{equation}  We have now set the stage to describe our derivation of negative specific heat.

Given the space available, we proceed to outline, rather than fully derive, our method of obtaining an explicit formula for the maximal entropy of this mean-field system
in the non-extensive thermodynamic limit (as defined above) from which we obtain an explicit, closed form formula for the specific heat.  A full description of the mathematics involved is forthcoming in another paper.

First it is important to note that our approach relies heavily on the steepest descent methods in \cite{Horwitz:1983} and spherical model approach in \cite{Berlin:1952, Hartman:1995, Keller:1960, Stanley:1968}.  The works of \cite{Lim:2006, Lim:2007} have preceded and inspired this work in their novel applications of the spherical model to barotropic vorticity models on the sphere.  These works lay the ground for our derivation.

The steepest-descent and consequently spherical model methods convert Dirac-delta functions into their Fourier-space equivalent integral representations.  The advantage of this integral representation is that one can
perform steepest-descent evaluation on it.  Therefore, the procedure for our derivation relies on
two closely related facts: the integral representation of the Dirac-delta function in a microcanonical distribution, for example:
\begin{equation}
\delta(Nx-Nx_0) = \int_{\beta_0-i\infty}^{\beta_0+i\infty} \frac{d\beta}{2\pi i} e^{N\beta(x-x_0)},
\end{equation} and the steepest-descent limit, again only an example:
\begin{equation}
\lim_{N\rightarrow\infty} \frac{1}{N} \int_{\beta_0-i\infty}^{\beta_0+i\infty} d\beta e^{N\beta(x-x_0)} = e^{\beta_0(x-x_0)},
\end{equation} which allows quantities such as entropy to be determined, provided we can determine what $\beta_0$ is, a method for determining which \cite{Horwitz:1983} provides.

Using the first fact, we can convert the microcanonical problem into the canonical problem by converting delta functions into integrals and re-ordering the phase-space ($\psi$) and parameter-space ($\beta$) integrals to arrive at
\begin{align}
 e^{S_N} = \int\frac{d\beta}{2\pi i}e^{\beta NH_0}Z_{bath},
\label{eqn:eS2}
\end{align} where
\begin{align}
 Z_{bath} = \int D\psi e^{-\beta E_N - \mu M_N}\delta(NR^2-M_N)
\label{eqn:zbath}
\end{align} is a canonical-like partition function for the system in an external heat-bath but with a microcanonical angular momentum constraint.  In the infinite $N$ limit, $Z_{bath}$ approaches the canonical partition function (i.e. the same function but without the Dirac-delta factor in the integrand of Equation \ref{eqn:zbath}), making their use interchangeable in the limit.

To do this, i.e. convert the microcanonical problem to a canonical one, one has to prove that the integrals are finite, and indeed we can but do not show it here.  One this is done, solving for the canonical partition function becomes the solution for the microcanonical partition function in steepest-descent, and, as is well-known, exponentials are much easier to integrate than Delta-functions over complicated energy surfaces.

The maximal entropy per filament per period, having the form,
\begin{equation}
 S_{max}(H_0) = \lim_{N\rightarrow\infty}N^{-1}S_N,
\end{equation} is where the limit comes into play.

One problem commonly seen with logarithmic, Coulomb-style interactions such as this is that as the number of ``particles'' (in this case vortex filaments) grows infinitely large, the interaction energy grows with the square of the number of filaments.  This type of super-extension is never seen in traditional molecular system where there is a clear distance cut-off, but it is seen in both Coulomb and gravitational systems like vorticity models and globular clusters.  The solution is to rescale the temperature, which in turn, rescales the interaction
energy.  Rescaling the temperature causes a chain of necessary scalings to restore the balance so that other quantities do not go to zero: $\beta'=\beta N$, $\alpha'=\alpha/N$, $p'=p/N$, and $H_0'=H_0/N$.  These are, of course, reasonable because only the interaction energy needs a rescaling.

With all these scalings there are no more 
mathematical obstructions,
and the maximal entropy can be found by the standard mathematical procedures in \cite{Horwitz:1983} and \cite{Berlin:1952}.  We provide only the final formula obtained in view of space constraints, but the procedure is quite 
straightforward
once the appropriate framework is set up:
\begin{align}
 S_{max}(H_0) = \beta_0' H_0' + \frac{\beta_0'}{4}\log(R^2) - \frac{1}{2\alpha\beta_0 R^2} - \beta_0'p'R^2.
\end{align} where
\begin{align}
 R^2 = \frac{\beta_0'^2\alpha' + \sqrt{\beta_0'^4\alpha'^2 + 32\alpha'\beta_0'^2p'}}{8\alpha'\beta_0'^2p'},
 \label{eqn:rsq}
\end{align} where the mean temperature, $\beta_0$, is as yet unknown.  This entropy is exact within the mean-field assumption for $N\rightarrow\infty$.

By \cite{Horwitz:1983}, we find the unknown multiplier, $\beta_0$, by relating the enthalpy per filament parameter, $H_0$, to the mean enthalpy, $NH_0 = \langle E_N+pM_N \rangle$, where $\langle \cdot\rangle$ denotes average against Equation \ref{eqn:gibbs}.

By Equation \ref{eqn:gibbs} the average enthalpy is given by
\begin{equation}
 \langle E_N + pM_N\rangle = \frac{\int D\psi E_N\delta(NH_0-E_N-pM_N)\delta(NR^2-M_N)}{\int D\psi \delta(NH_0-E_N-pM_N)\delta(NR^2-M_N)},
\end{equation} and, again going through some steepest-descent-based calculations given in \cite{Horwitz:1983}, we find the formula,
\begin{align}
 H_0' &= \frac{\partial}{\partial\beta_0'}\left(-\frac{\beta_0'}{4}\log R^2 + \frac{1}{2\alpha'\beta_0' R^2} + \beta_0'p'R^2\right)
 \label{eqn:energyprime}
\end{align} exactly.  We cannot give an explicit expression for $\beta_0$ because it is a root of a transcendental equation, but such is unnecessary for the following negative specific heat result:

We define specific heat at constant generalized pressure, $p$,
\begin{equation}
 c_p = -\beta_0^2\frac{\partial H_0}{\partial \beta_0}
\end{equation} and after evaluating with Equation \ref{eqn:energyprime} and simplifying (dropping primes and $0$-subscripts)
\begin{align}
 c_p = \frac{\beta}{4}\left(\frac{\alpha\beta^2}{\sqrt{\alpha\beta^2(\alpha\beta^2+32p)}}-1\right).
 \label{eqn:spheat}
\end{align}

\begin{figure}
\begin{center}
\includegraphics[width = 0.48\textwidth]{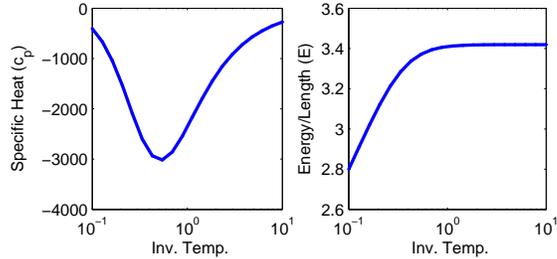}
\end{center}
\caption{The specific heat at constant pressure (Equation \ref{eqn:spheat}) for the thermally isolated system is negative, meaning that the constant pressure Enthalpy Per Length (Equation \ref{eqn:energyprime}) decreases with increasing temperature.  (Here $\alpha'=5\times 10^5$ and $p'=8\times 10^4$.)}
\label{fig:spheat} 
\end{figure}

\begin{figure}
\begin{center}
\includegraphics[width = 0.3\textwidth]{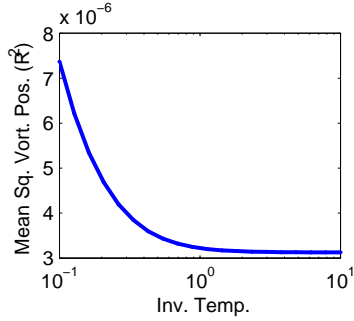}
\end{center}
\caption{The Mean Square Vortex Position (Equation \ref{eqn:rsq}) increases exponentially at high temperature, while it is nearly constant at low-temperature. (Here $\alpha=5\times 10^5$ and $p=8\times 10^4$.)}
\label{fig:rsq}
\end{figure}

Equation \ref{eqn:spheat} is significant.  It indicates that the specific heat is not only negative
for this system, but \emph{strictly} negative if parameters are non-zero (Figure \ref{fig:spheat}).  In the low-temperature (large $\beta$) case, for constant field strength, $R^2$ does not change significantly with temperature indicating that filaments are in a stable configuration for a large range of low-temperatures.  Because the filaments do not move
relative to one another at low-temperatures and the self-induction is negligible, the enthalpy does not change.  When temperature
becomes high the internal entropy causes a massive expansion in the overall size of the system (Figure \ref{fig:rsq}), and
energy of the logarithmic interaction decreases far more than the enthalpy
of the self-induction increases.  The strong magnetic field absorbs this energy, but, since
it is assumed to be an infinitely massive reservoir able to maintain the enthalpy at $H_0$, the confinement remains constant.

On a side note, because the specific heat in Equation \ref{eqn:spheat} does not cross the axis (i.e. is never positive) for any positive parameters, the expansion
in $R^2$ in the canonical system is not likely a phase transition in $\beta$, but a continuous ``transition'' \cite{Lynden:1977}.  The system appears to change behavior significantly, but it changes without any discontinuity in the free energy or specific heat.

That only a local-induction approximation is necessary for this analysis is interesting because it is so simple and so moderately different from 2D vortex system which do not exhibit negative specific heat in statistical equilibrium.  What happens in the nearly parallel vortex filament case can be explained in steps similar to the catastrophic gravo-thermal collapse: (1) a vortex's Brownian motion causes it or part of it to move away from the center, (2) potential energy decreases, (3) the vortices in the center can move closer together and temperature increases.

As mentioned in the preceding paragraph, the negative specific heat indicates a runaway reaction (i.e. the fixed energy, fixed angular momentum equilibrium point is meta-stable) in the microcanonical system that models strongly confined, single-sign generalized vortex filaments.  This is the insight we have from gravo-thermal collapse:  We hypothesize that this kind of meta-stability, observed in gravo-thermal collapse of globular clusters \cite{Lynden:1968}, can lead to two possible outcomes: (1) a collapse similar to globular clusters in which an outer halo of columns separates from an inner core that collapses in on itself, possibly resulting in nuclear fusion, (2) a turbulent expansion of the entire system.  Further research will focus on answering this question, but clearly 3D effects, even only a LIA, are crucial.

{\bf Acknowledgments}:
This work is supported by ARO grant W911NF-05-1-0001 and DOE grant 
DE-FG02-04ER25616. 
\pagebreak
\bibliography{pimcposter}
\end{document}